\documentclass[aps,prd,twocolumn,superscriptaddress,floatfix]{revtex4-2}
\usepackage{graphicx}
\usepackage{amstext} 
\usepackage{array}   
\newcolumntype{L}{>{$}l<{$}} 
\usepackage{pstricks}
\usepackage{color}
\usepackage{placeins}

\begin{document}

\title{Triple top quark production in Standard Model}

\author{E.~Boos}
\author{L.~Dudko}
\affiliation{Skobeltsyn Institute of Nuclear Physics, Lomonosov Moscow State University, Russia, Moscow}



\begin{abstract}
The triple top quark production processes have been investigated and calculated in the scope of the Standard Model. The cross sections for different channels are provided for the proton-proton collision energies of $\sqrt{s}=$ 14 and 100 TeV. The importance of the electroweak contribution has been demonstrated. For the main channels the interference between the gluon and the weak bosons mediated contributions is negative and significant, therefore the complete set of diagrams has to be taken into account. Estimated total rates are about 1.9 fb for $\sqrt{s}=14$ TeV and 530 fb for $\sqrt{s}=100$ TeV. 
A simple estimation of the uncertainty of the calculated cross sections gives about 20\%. The integrated luminosity of 3 ab$^{-1}$ at HL-LHC allows expecting about 5700 events, giving a chance to detect this rare process. 
\end{abstract}


\maketitle

\section{Introduction}
Top quark was discovered in pair top anti-top quark production~\cite{D0:1995jca,Abe:1995hr}, which have the highest cross section. The single production of the top quark occurs due to the electroweak interaction caused by the vertex tWb. The single top quark production was also discovered at the Tevatron  experiments~\cite{Aaltonen:2009jj,Abazov:2009ii}. The associated production of the top quark with W~\cite{Chatrchyan:2014tua,Aad:2015eto}, Z~\cite{Sirunyan:2018zgs,Chatrchyan:2013qca,Aad:2015eua} or Higgs~\cite{Sirunyan:2018hoz,Aaboud:2018urx} bosons were first observed at the LHC experiments. The present level of the experimental sensitivity is close to observe the four top quark production~\cite{Aaboud:2018jsj,Sirunyan:2019nxl,Sirunyan:2019wxt}. In the Standard Model (SM) there are no production processes with exactly three top quarks, but there are several processes of the three top quark production in association with other particles. The SM total cross sections of the triple top quark production processes have been calculated previously~\cite{Barger:2010uw,Chen:2014ewl,Malekhosseini:2018fgp}.
The triple top quark production can be sensitive to a number of higher dimensional operators leading to flavor changing neutral currents involving the top quark~\cite{Khanpour:2019qnw,Cao:2019qrb} and production of new resonances~\cite{Iguro:2017ysu,Kohda:2017fkn,Cho:2019stk}.
Search for any beyond the SM (BSM) manifestation in the triple top quark production requires accurate  computations of the SM contribution. The main aim of our study is to calculate the electroweak and QCD SM contributions, as well as interference between them, of all dominating  triple top quark production processes in association with a lighter quark including the b-quark (p,p$\to$ t,t($\bar{\rm t}$),$\bar{\rm t}$,q) or with the W boson (p,p$\to$ t,t($\bar{\rm t}$),$\bar{\rm t}$,W). All the computations are performed at LO  by means of the CompHEP package~\cite{Boos:2004kh,Pukhov:1999gg} which allows getting symbolic and numerical results separately for different subsets of Feynman diagrams and, also, the interference terms. Numerical results are given for proton-proton collisions with energies $\sqrt{s}=14$ and $100$ TeV, using the PDF set NNPDF23-nlo-as-0118~\cite{Ball:2012cx} as it is realized in LHAPDF5.9.1~\cite{Buckley:2014ana}, factorization and renormalization scales are chosen as $\rm Q=3M_{top}/2$.
The paper is organizes as follows. 
   In the next section, we briefly discuss gauge invariant classes of contributing leading order $2 \to 4$ diagrams. In the next three sections some details of computations and numerical results for the cross sections are given for three main channels  $\rm pp \to \rm t \,t\,\bar{t}\,q$ / $\rm t\,\bar{\rm t}\,\bar{\rm t}\,q$, q=u,d,c,s;
$\rm pp \to \rm t\,t\,\bar{\rm t}\,{\rm W^-}$ / $\rm t\,\bar{\rm t}\,\bar{\rm t}\,{\rm W^+}$; 
and $\rm pp \to \rm t\,t\,\bar{\rm t}\,\bar{\rm b}$ / $\rm t\,\bar{\rm t}\,\bar{\rm t}\,{\rm b}$ respectively. The total cross section for the triple top quark production process is presented in the last section, followed by a short conclusion.  

\section{Gauge invariant subsets of Feynman diagrams}
The complete set of Feynman diagrams for any triple top quark production consists of several parts. As an example, let as consider the leading subprocess  u,b$\to$ t,t,$\bar{\rm t}$,d. There are two gauge invariant subsets~\cite{Boos:1999qc} of diagrams mediated by gluons and electroweak bosons. The Feynman diagrams for gluon mediated contribution are shown in Fig.~\ref{Gdiags}. The diagrams are similar to the t-channel production of a single top quark with additional splitting of a gluon into a pair of top quarks.  
The fourteen diagrams for the electroweak mediated contribution are shown in Fig.~\ref{noGdiags}. One can see the thirteen diagrams with virtual photon, Z or H boson decaying to top pair and the last t-channel diagram number 14. Since the diagrams with virtual photon, Z and H bosons are topologically the same as gluon mediated diagrams, one could expect that their contribution is much smaller and can be neglected in comparison with gluon mediated contribution. However, if we take four gluon mediated diagrams and  add diagram 14 (Fig.~\ref{noGdiags}) the rate will be about ten times larger than the correct one.
 The calculation of gluon mediated set of diagrams (Fig.~\ref{Gdiags}) gives the cross section of 0.10 fb at $\sqrt{s}=14$, in unitary gauge. The same result is obtained in t\'Hooft-Feynman gauge, as it should be, for the gauge invariant subset of diagrams. In case we add diagram 14 of Fig.~\ref{noGdiags} to the gluon mediated diagrams, the cross section rises in ten times to 1.4 fb (in unitary gauge) reflecting the violation of gauge invariance. The complete weak bosons mediated set of diagrams shown in Fig.~\ref{noGdiags} gives the cross section 0.12 fb, which is of the same order as the gluon mediated set of diagrams. The gauge invariance of the subclasses listed in Fig.~\ref{Gdiags} and Fig.~\ref{noGdiags} has been checked with the calculation in both unitary and t\'Hooft-Feynman gauges leading to the same results. The interference between these two sets of Feynman diagrams in Figs.~\ref{Gdiags},~\ref{noGdiags} is also gauge invariant and has to be taken into account. Calculation of the complete set of all diagrams in Figs.~\ref{Gdiags},~\ref{noGdiags}, including interference between them, provides the total cross section of 0.12 fb. The total cross section is a sum of three gauge invariant contributions:   0.12 fb (tot) = 0.10 fb (gluon) + 0.12 fb (electroweak) - 0.10 fb (interference). One can see how significant the negative interference is. Therefore, such an example demonstrates that the complete set of Feynman diagrams has to be calculated in order to get the correct cross section of the triple top quark production. All numbers above are given with the requirement on transverse momenta of d-quark $P_T^d > 10$ GeV.  
\begin{widetext}
\begin{figure*}
\includegraphics[width=0.75\textwidth,clip]{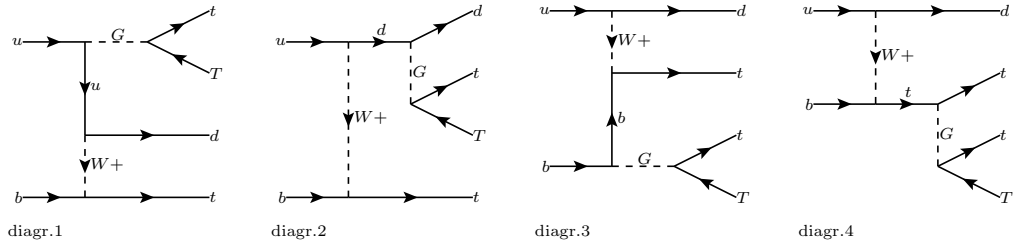}%
\caption{\label{Gdiags} Feynman diagrams for the gluon mediated contribution to u,b$\to$ t,t,$\bar{\rm t}$,d process.}
\end{figure*}
\begin{figure*}
\includegraphics{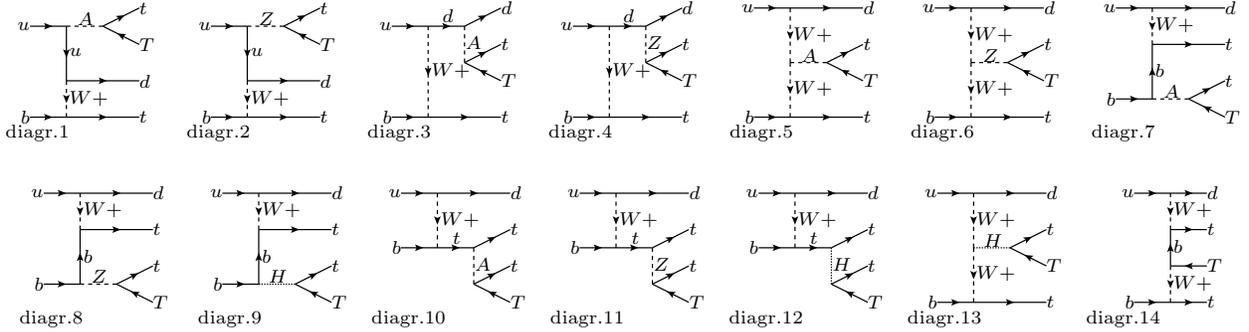}%
\caption{\label{noGdiags} Feynman diagrams for the weak bosons mediated contribution to u,b$\to$ t,t,$\bar{\rm t}$,d process.}
\end{figure*}
\end{widetext}

\section{\label{tttq} Processes with $\rm tt\bar{t}q$ and $\rm t\bar{\rm t}\bar{\rm t}q$ final states}
One of the main contribution comes from the associated triple top quark production with the light flavor quark. There are eight representative processes p,p$\to$ t,t,$\bar{\rm t}$,q  listed in Table~\ref{tab:ttTq} and eight representative processes p,p$\to$ t,$\bar{\rm t}$,$\bar{\rm t}$,q listed in Table~\ref{tab:tTTq}. As in the previous section, the requirement  $P_T^q > 10$ GeV is used. In the tables, the cross sections for specific initial states are given as shown, but for the sums and total cross sections, an additional factor of two is included, reflecting the permutations of two partons in the initial states. As one can see from the tables, the rate of the process with signature t,t,$\bar{\rm t}$,q is about 2.6 times larger than for t,$\bar{\rm t}$,$\bar{\rm t}$,q signature due to the difference in parton distribution functions (PDF) of up-type and down-type quarks. This is similar to the difference in rates of t-channel single top and single anti-top quark production. The leading contribution comes from the first generation quarks in the initial states, subproceses 1 in Table~\ref{tab:ttTq} and in Table~\ref{tab:tTTq}. The subprocesses 2, 4, 5 in Table~\ref{tab:ttTq} and subprocesses 2, 3, 4 in Table~\ref{tab:tTTq} are suppressed by PDF. The subprocesses 3 in Table~\ref{tab:ttTq} and 5 in Table~\ref{tab:tTTq} are suppressed by Cabibbo–Kobayashi–Maskawa (CKM) matrix elements. The subprocesses 6, 7, 8 in Table~\ref{tab:ttTq} and Table~\ref{tab:tTTq} are suppressed  by both PDF and CKM. The contribution of the suppressed processes  is about 13\% to the total rate of all processes listed in Table~\ref{tab:ttTq} and about 29\% of all processes listed in Table~\ref{tab:tTTq}. The contributions of s and $\rm\bar{s}$ in the initial states (subprocesses 4 in Table~\ref{tab:ttTq} and subprocess 3 in Table~\ref{tab:tTTq}) are slightly different due to the difference of s and $\rm\bar{s}$ PDF in NNPDF2.3~\cite{Ball:2013gsa}.
The total rate is about 0.38 fb allowing to expect more than a thousand events for High Luminosity LHC with 3 ab$^{-1}$ of integrated luminosity.
\begin{table}[h!]
		\begin{center}
		\caption{Cross sections for individual contributing processes and total cross sections for the p,p$\to$ t,t,$\bar{\rm t}$,q production. Permutations of the initial partons are included to the total cross sections. The $P_T^q > 10$ GeV cut was applied.}
		\label{tab:ttTq}
		\begin{tabular}{|c|l|c|}
			\hline
\rm tt$\bar{\rm t}$q & \text{subprocess} & \rm Cross section [pb] \\ \hline
\rule{0pt}{1em}1 & u, b $\to$ d, t, t, $\bar{\rm t}$ &   1.19e-04  \\ 
 2 & $\bar{\rm d}$, b $\to$ $\bar{\rm u}$, t, t, $\bar{\rm t}$ &   6.45e-06 \\ 
 3 & u, b $\to$ s, t, t, $\bar{\rm t}$ &   6.22e-06  \\ 
 4 & $\bar{\rm s}$, b $\to$ $\bar{\rm c}$, t, t, $\bar{\rm t}$ &   2.69e-06  \\ 
 5 & c, b $\to$ s, t, t, $\bar{\rm t}$ &   2.60e-06   \\ 
 6 & $\bar{\rm d}$, b $\to$ $\bar{\rm c}$, t, t, $\bar{\rm t}$ &   3.37e-07  \\ 
 7 & $\bar{\rm s}$, b $\to$ $\bar{\rm u}$, t, t, $\bar{\rm t}$ &   1.41e-07  \\ 
 8 & c, b $\to$ d, t, t, $\bar{\rm t}$ &   1.36e-07  \\ 
\hline
G  & Sum of gluon diags. & 2.28e-04   \\
EW & Sum of EW diags. & 2.88e-04   \\
Int & Interference & -2.41e-04  \\ \hline
    & Total cross section  & 2.75e-04 \\ \hline
		\end{tabular}
		\end{center}
 \end{table}

 \begin{table}[h!]
		\begin{center}
		\caption{Cross sections for individual contributing processes and total cross sections for the p,p$\to$ t,$\bar{\rm t}$,$\bar{\rm t}$,q production. Permutations of the initial partons are included to the total cross sections. The $P_T^q > 10$ GeV cut was applied.}
		\label{tab:tTTq}
		\begin{tabular}{|c|l|c|}
			\hline
\rm t$\bar{\rm t}$$\bar{\rm t}$q & \text{subprocess} & \rm Cross section [pb]  \\ \hline
\rule{0pt}{1em}1 & d, $\bar{\rm b}$ $\to$ u, t, $\bar{\rm t}$, $\bar{\rm t}$ &   3.92e-05   \\ 
 2 & $\bar{\rm u}$, $\bar{\rm b}$ $\to$ $\bar{\rm d}$, t, $\bar{\rm t}$, $\bar{\rm t}$ &   7.33e-06   \\ 
 3 & s, $\bar{\rm b}$ $\to$ c, t, $\bar{\rm t}$, $\bar{\rm t}$ &   3.20e-06   \\ 
 4 & $\bar{\rm c}$, $\bar{\rm b}$ $\to$ $\bar{\rm s}$, t, $\bar{\rm t}$, $\bar{\rm t}$ &   2.60e-06   \\ 
 5 & d, $\bar{\rm b}$ $\to$ c, t, $\bar{\rm t}$, $\bar{\rm t}$ &   2.05e-06   \\ 
 6 & $\bar{\rm u}$, $\bar{\rm b}$ $\to$ $\bar{\rm s}$, t, $\bar{\rm t}$, $\bar{\rm t}$ &   3.83e-07    \\ 
 7 & s, $\bar{\rm b}$ $\to$ u, t, $\bar{\rm t}$, $\bar{\rm t}$ &   1.67e-07   \\ 
 8 & $\bar{\rm c}$, $\bar{\rm b}$ $\to$ $\bar{\rm d}$, t, $\bar{\rm t}$, $\bar{\rm t}$ &   1.36e-07   \\ 
\hline
G  & Sum of gluon diags. & 1.05e-04  \\
EW &Sum of EW diags.& 1.20e-04   \\
Int & Interference & -1.15e-04  \\ \hline
    &Total cross section  & 1.10e-04   \\ \hline
		\end{tabular}
		\end{center}
 \end{table}

\section{\label{tttW} Processes with $\rm tt\bar{\rm t}{\rm W^-}$ and $\rm t\bar{\rm t}\bar{\rm t}{\rm W^+}$ final states}
Consider the triple top quark production in association with W boson. 
There are 59 $2\to4$ contributing Feynman diagrams with anti-b-quark in the initial state with W$^+$ in the final state and the same number of diagrams for the W$^-$ in final state with b-quark in the initial state. Since the b-quark and anti-b-quark PDF are the same in the proton, the cross sections are equal for charge symmetric final states with W$^+$ and W$^-$. As in previous example, 59 diagrams are splitted to two gauge invariant subclasses, gluon and electroweak mediated parts. In the Table~\ref{tab:tttW}, the cross sections for specific initial states are given, and for the sums and total cross sections, an additional factors of two for W$^+$ and W$^-$ are included, reflecting the permutations of two partons in the initial states. The contribution of electroweak diagrams is of the same order as a gluon mediated contribution, and the interference between them is significant and negative. 
All calculations are done in so called 5-flavour scheme 
(5FS)~\cite{Collins:1998rz,Kramer:2000hn,Rainwater:2002hm,Maltoni:2003pn,Boos:2003yi,Harlander:2003ai,Moretti:2003px,Dittmaier:2003ej,Campbell:2004pu,Dawson:2005vi,Maltoni:2012pa,Wiesemann:2014ioa,Harlander:2015xur,Spira:2016ztx,Duhr:2020kzd} with a b-quark in the initial state as a parton in colliding proton. Similar to the case of associated tW-channel single top quark production in addition to the process g,b$(\bar{\rm b}) \to \rm t,t(\bar{\rm t}),\bar{\rm t},{\rm W^-(W^+)}$ there is a process with two gluons in the initial and an extra b-quark in the final state g,g$\to \rm t,t(\bar{\rm t}),\bar{\rm t},{\rm W^-(W^+)},\bar{\rm b}(\rm b)$. The later process is a part of next-to-leading order (NLO) tree level corrections in 5FS.  This part interferes with four top quark production including subsequent top quark decay and has to be taken into account for calculations at NLO level and especially for event simulation~\cite{Frixione:2008yi,Denner:2010jp,Bevilacqua:2010qb,Cascioli:2013wga,Jezo:2016ujg}.
\begin{table}[h!]
		\begin{center}
		\caption{Cross sections for individual contributing processes and total cross sections for the p,p$\to$ t,t($\bar{\rm t}$),$\bar{\rm t}$,W production. Permutations of the initial partons are included to the total cross sections.}
		\label{tab:tttW}
		\begin{tabular}{|c|l|c|}
			\hline
\rm tt($\bar{\rm t}$)$\bar{\rm t}$W & \text{subprocess} & \rm Cross section [pb]  \\ \hline
\rule{0pt}{1em}1 & $\bar{\rm b}$, g $\to$ $\rm W^+$, t, $\bar{\rm t}$, $\bar{\rm t}$ &   3.40e-04  \\ 
2 & b, g $\to$ $\rm W^-$, t, t, $\bar{\rm t}$ &   3.40e-04    \\ 
\hline
G   &Sum of gluon diags. & 1.01e-03   \\
EW &Sum of EW diags. & 0.962e-03  \\
Int & Interference & -0.612e-03  \\ \hline
    & Total cross section  & 1.36e-03   \\ \hline
		\end{tabular}
		\end{center}
 \end{table}

\section{\label{tttb} Processes with $\rm tt\bar{\rm t}\bar{\rm b}$ and $\rm t\bar{\rm t}\bar{\rm t}{\rm b}$ final states}
Additional small contribution to triple top quark production comes from the associated production with b-quark. The diagrams correspond to s-channel single top quark production with additional top quarks coming from virtual gluon, photon, Higgs, Z- or W-bosons. 
There are four representative processes p,p$\to$ t,t,$\bar{\rm t},\bar{\rm b}$ with $\bar{\rm b}$-quark listed in Table~\ref{tab:ttTb}  and four representative processes p,p$\to$ t,$\bar{\rm t}$,$\bar{\rm t}$,b  with b-quark listed in Table~\ref{tab:tTTb}. As for the light quark in Sec.~\ref{tttq} the requirement  $P_T^b > 10$ GeV is used. 
Similar to the tables in Sec.~\ref{tttq} in Tables~\ref{tab:ttTb},~\ref{tab:tTTb} the cross sections for specific initial states are given as shown, but for the sums and total cross sections, an additional factor of two is included, reflecting the permutations of two partons in the initial states.
As can be seen from the tables the main contribution comes from the processes with $\rm u,\bar d$ and $\rm d,\bar u$ initial states. The first contribution is about 2.4 times larger than the second due to the differences in PDF.
The electroweak part (EW) is roughly 3\% of the gluon mediated contribution (G). The interference (Int) is positive and of the same order as the electroweak contribution.  
\begin{table}[h!]
		\begin{center}
		\caption{Cross sections for individual contributing processes and total cross sections for the p,p$\to$ t,t,$\bar{\rm t}$,$\bar{\rm b}$ production. Permutations of the initial partons are included to the total cross sections. The $P_T^{\rm \bar b} > 10$ GeV cut was applied.}
		\label{tab:ttTb}
		\begin{tabular}{|c|l|c|}
			\hline
\rm tt$\bar{\rm t}$$\bar{\rm b}$ & \text{subprocess} & \rm Cross section [pb]  \\ \hline
\rule{0pt}{1em}1 & u, $\bar{\rm d}$ $\to$ t, t, $\bar{\rm t}$, $\bar{\rm b}$ &   5.35e-05    \\ 
 2 & u, $\bar{\rm s}$ $\to$ t, t, $\bar{\rm t}$, $\bar{\rm b}$ &   1.57e-06    \\ 
 3 & c, $\bar{\rm s}$ $\to$ t, t, $\bar{\rm t}$, $\bar{\rm b}$ &   9.91e-07      \\ 
 4 & c, $\bar{\rm d}$ $\to$ t, t, $\bar{\rm t}$, $\bar{\rm b}$ &   1.14e-07    \\ 
\hline
G & Sum of gluon diags. & 1.06e-04  \\
EW & Sum of EW diags. & 3.43e-06 \\
Int & Interference & 2.57e-06  \\ \hline
 & Total cross section & 1.12e-04   \\ \hline
		\end{tabular}
		\end{center}
 \end{table}
\begin{table}[h!]
		\begin{center}
		\caption{Cross sections for individual contributing processes and total cross sections for the p,p$\to$ t,$\bar{\rm t}$,$\bar{\rm t}$,b production. Permutations of the initial partons are included to the total cross sections. The $P_T^b > 10$ GeV cut was applied.}
		\label{tab:tTTb}
		\begin{tabular}{|c|l|c|}
			\hline
\rm t$\bar{\rm t}$$\bar{\rm t}$b & \text{subprocess} & \rm Cross section [pb] \\ \hline
 \rule{0pt}{1em}1 & d, $\bar{\rm u}$ $\to$ t, $\bar{\rm t}$, $\bar{\rm t}$, b &   2.25e-05     \\ 
 2 & s, $\bar{\rm c}$ $\to$ t, $\bar{\rm t}$, $\bar{\rm t}$, b &   1.13e-06      \\ 3 & d, $\bar{\rm c}$ $\to$ t, $\bar{\rm t}$, $\bar{\rm t}$, b &   5.51e-07    \\ 
 4 & $\bar{\rm u}$, s $\to$ t, $\bar{\rm t}$, $\bar{\rm t}$, b &   1.47e-07     \\ 
\hline
G  & Sum of gluon diags. & 4.60e-05  \\
EW & Sum of EW diags. & 1.44e-06  \\
Int & Interference & 1.16e-06  \\ \hline
 & Total cross section & 4.86e-05   \\ \hline
		\end{tabular}
		\end{center}
 \end{table}
 
\section{Total cross sections and uncertainties}
The total cross sections for the triple top quark production in different channels are summarized in Table~\ref{tab:tot14} for the proton-proton collisions with $\sqrt{s}=14$ TeV. The total rate of 1.9 fb allows to expect 5700 events for 3 ab$^{-1}$ integrated luminosity giving a chance to detect the process at HL-LHC. If one excludes all-hadronic and $\tau$+jets channels (branching ratio is about 0.6) and estimates a total experimental acceptance as 10\% one can expect about 300 detected events.
In Table~\ref{tab:tot100} the cross sections for the proton-proton collisions at FCC energy $\sqrt{s}=100$ TeV are shown. In this case the same requirement for momentum transverse of light- or b-quark $P_T^{\rm q,b} > 10$ GeV is used. Note the significant increase of the total cross section. The main reason for this is a rapid growing of the gluon parton density g(x) in protons with a decrease of the momentum fraction x. The uncertainties of the calculations are estimated as follows. The factorisation and renormalisation scales uncertainty is taken as $\rm\Delta\sigma_Q=|\sigma(Q=3M_{\rm top})-\sigma(Q=3M_{\rm top}/4)|/2$, and $\rm \delta\sigma_Q=\Delta\sigma_Q/\sigma(Q=3M_{\rm top}/2)$. The PDF uncertainty is estimated as $\rm \Delta\sigma_{PDF}=|\sigma(CTEQ6l1)-\sigma(NNPDF23)|$, and $\rm \delta\sigma_{PDF}=\Delta\sigma_{PDF}/\sigma(NNPDF23)$. The estimation of the $\rm \delta\sigma_{PDF}$ is simple and does not follow the PDF4LHC recipe~\cite{Butterworth:2015oua} but it gives qualitatively the order of the PDF uncertainty. As one can see from the Tables~\ref{tab:tot14},\ref{tab:tot100} the scale uncertainty  $\rm \delta\sigma_Q$ is about two or three times higher than the PDF uncertainty $\rm \delta\sigma_{PDF}$.
\begin{table}[h!]
		\begin{center}
		\caption{Total cross sections of different contributions to triple top quark production in pp collisions at $\sqrt{s}=14$ TeV.}
		\label{tab:tot14}
		\begin{tabular}{|c|c|c|c|}
			\hline
Process & Cross sec. [pb] & $\delta\sigma_{Q}$, \%  & $\delta\sigma_{\rm PDF}$, \%\\ \hline
\rule{0pt}{1em}p, p $\to$ W$^-$,  t,  t ,  $\bar{\rm t}$               & 6.8e-04  & 13 & 6 \\
p, p $\to$ W$^+$,  t,  $\bar{\rm t}$,  $\bar{\rm t}$    & 6.8e-04  & 13 & 6  \\
p, p $\to$ q$^\prime$, t, t, $\bar{\rm t}$              & 2.7e-04 &  12 & 14\\
p, p $\to$ q$^\prime$, t, $\bar{\rm t}$, $\bar{\rm t}$  & 1.1e-04  &  13 & 4 \\
p,  p $\to$ $\bar{\rm b}$,  t,  t,  $\bar{\rm t}$       & 1.1e-04  & 35 & 13\\
p,  p $\to$ b,  t,  $\bar{\rm t}$,  $\bar{\rm t}$       & 4.9e-05  & 35 & 4\\ \hline
Total p,  p $\to$ X,  t,  t ($\bar{\rm t}$),  $\bar{\rm t}$ & 1.9e-03 & 15 &  7 \\ \hline
		\end{tabular}
		\end{center}
 \end{table}
\begin{table}[h!]
		\begin{center}
		\caption{Total cross sections of different contributions to triple top quark production in pp collisions at $\sqrt{s}=100$ TeV.}
		\label{tab:tot100}
		\begin{tabular}{|c|c|c|c|}
			\hline
Process & Cross sec. [pb] & $\delta\sigma_{Q}$, \%  & $\delta\sigma_{\rm PDF}$, \%\\ \hline
\rule{0pt}{1em}p, p $\to$ W$^-$,  t,  t ,  $\bar{\rm t}$               & 2.4e-01  & 15 & 4 \\
p, p $\to$ W$^+$,  t,  $\bar{\rm t}$,  $\bar{\rm t}$    & 2.4e-01 &  15 & 4  \\
p, p $\to$ q$^\prime$, t, t, $\bar{\rm t}$              & 3.1e-02 &  4 & 7\\
p, p $\to$ q$^\prime$, t, $\bar{\rm t}$, $\bar{\rm t}$  & 1.8e-02 &  4 & 4 \\
p,  p $\to$ $\bar{\rm b}$,  t,  t,  $\bar{\rm t}$       & 2.6e-03  & 12 & 4\\
p,  p $\to$ b,  t,  $\bar{\rm t}$,  $\bar{\rm t}$       & 1.7e-03  & 12 & 4\\ \hline
Total p,  p $\to$ X,  t,  t ($\bar{\rm t}$),  $\bar{\rm t}$ & 5.3e-01 & 14 & 4 \\ \hline
		\end{tabular}
		\end{center}
 \end{table}
 
\section{Conclusion}
Different channels of triple top quark production are considered in the scope of the SM.
The performed calculations and the provided results demonstrate the importance of the electroweak contribution and a significant negative interference between  gluon  and weak boson mediated contributions. The total cross section of the triple top quark production is at the level of 1.9 fb for the proton-proton collisions at $\sqrt{s}=14$ TeV and 530 fb at $\sqrt{s}=100$ TeV. The calculated total cross sections at $\sqrt{s}=14$ TeV are in a good agreement with previous independent calculations using the MadGraph package~\cite{Barger:2010uw,Chen:2014ewl,Malekhosseini:2018fgp}. The cross section at $\sqrt{s}=14$ TeV is rather small although its level seems to be enough to detect the triple top quark production taken into account expected luminosity 3 ab$^{-1}$ at HL-LHC. From the other hand, the small SM rate provides a possibility to search for beyond of SM (BSM) contributions which may increase the cross section. However, dedicated calculations are needed to study concrete BSM manifestation.  

\begin{acknowledgments}
This research has been supported by the Interdisciplinary Scientific and Educational School of Moscow State University ``Fundamental and Applied Space Research''.
\end{acknowledgments}
\FloatBarrier

\bibliography{triple_top}

\end{document}